\documentclass[prl,aps,twocolumn, floatfix, superscriptaddress]{revtex4}
\usepackage{amsmath,bm,graphicx}
\usepackage{amssymb}
\usepackage{amsfonts}
\usepackage[dvips,usenames]{color}
\usepackage{natbib}
\newcommand{\PM}[1]{{\color{black}{{#1}}}}

\begin{document}

\title{Entropically-induced asymmetric passage times of charged tracers across corrugated channels}

\author{Paolo Malgaretti}
\email[Corresponding Author : ]{malgaretti@is.mpg.de}
\affiliation{Max-Planck-Institut f\"{u}r Intelligente Systeme, Heisenbergstr. 3, D-70569
Stuttgart, Germany}
\affiliation{IV. Institut f\"ur Theoretische Physik, Universit\"{a}t Stuttgart,
Pfaffenwaldring 57, D-70569 Stuttgart, Germany}
\affiliation{Department de Fisica Fonamental, Universitat de Barcelona, Spain}

\author{Ignacio Pagonabarraga}
\affiliation{Department de Fisica Fonamental, Universitat de Barcelona, Spain}
\author{J. Miguel Rubi}
\affiliation{Department de Fisica Fonamental, Universitat de Barcelona, Spain}
\date{\today}

\begin{abstract}
We analyze the diffusion of charged and neutral tracers suspended in an electrolyte embedded in a \PM{channel of varying cross-section}. Making use of systematic approximations, the diffusion equation governing the motion of tracers  is mapped into an effective $1D$ equation describing the dynamics along the longitudinal axis of the channel where its varying-section  is encoded as an effective entropic potential. This simplified approach allows us to characterize tracer diffusion under generic confinement  by measuring their mean first passage time (MFPT). In particular, we show that the interplay between  geometrical confinement and  electrostatic interactions strongly affect the MFTP of tracers across corrugated channels hence leading to alternative means to control tracers translocation across charged pores. Finally, \PM{our results show that} the MFPTs of a charged tracer in opposite directions along an asymmetric channel may differ
We expect our results to be relevant for biological as well synthetic devices whose dynamics is controlled by the detection of diluted tracers.
\end{abstract}


\maketitle
\section{I Introduction}

The control of the properties of tracer transport along channels or pores is a key issue for a variety of situations. For example, the development of micro- and nano- fluidic devices relies on the understanding of the tracer motion in an electrolyte embedded in micro- and nano- metric confinement~\cite{Lyderic_Charlaix}. Moreover, several biological processes such as neuronal signaling, ion pumping, photosynthesis and ATPase, just to mention a few, rely on the transport of ions across membranes or through channels~\cite{Albers}. 
In general, the structure of the  conduit tracers travel through is  inhomogeneous and it may present bottlenecks or cavities that can alter the overall transport properties, as it has been recently shown for both charged~\cite{Ghosal,Siwy,hanggi,Lyderic_nature,Umberto,letter-electrokin,paolo_macromolecules} as well neutral systems~\cite{Reguera2006,Dagdug,paolo_jcp_2013,Umberto2015}. In particular, inhomogeneities in the properties of the channel can lead to rectification~\cite{hanggi}, diode-like behavior~\cite{Berezhkovskii2009,Lyderic_PRL}, or recirculation and negative mobility~\cite{letter-electrokin,paolo_macromolecules}. 
In many of the aforementioned scenarios the main aim is to control the current of tracers at steady state. However, diverse biological as well as synthetic scenarios are controlled by the recognition of very diluted receptors as it is for diffusion limited reactions and pattern forming systems~\cite{RiceBook}, transport across nuclear~\cite{Licata2009,Kutay1999} or plasma~\cite{calero} membrane or, in general, for detectors of solutes diffusing through porous media~\cite{zeolites,Willmott2015}. \PM{Concerning the latter, cone-like pores have been particularly exploited in resistive-pulse sensing techniques to measure properties of diverse particle raging from micro- to nano-metric scales~\cite{Saleh2003,Ito2004,Heins2005,Arjmandi2012}}.For such systems the time a tracer takes to reach a given target for the first time, namely the Mean First Passage Time (MFPT), constitutes a standard and useful indicator.

\begin{figure}
\includegraphics[scale=0.25]{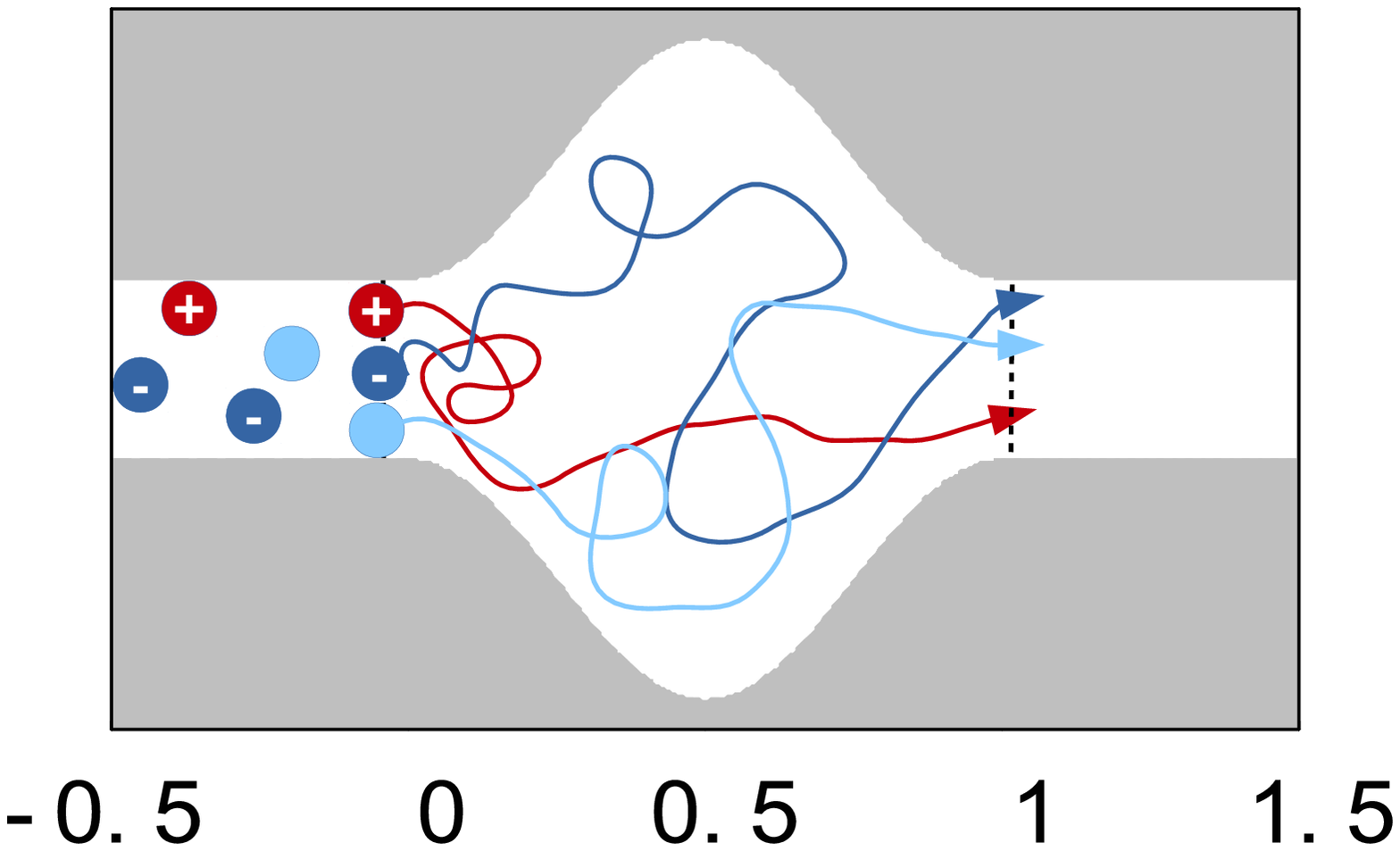}\includegraphics[scale=0.25]{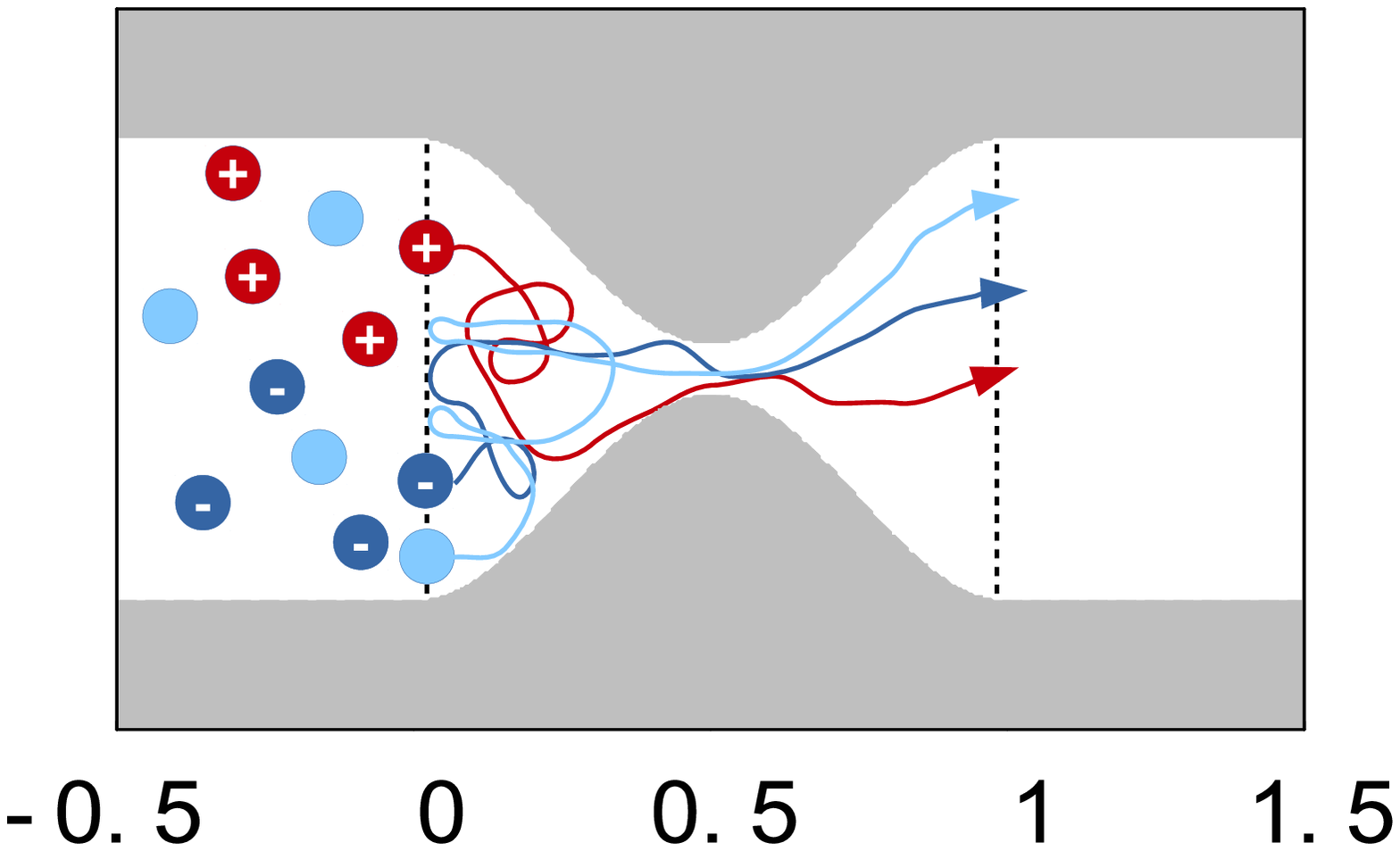}
 \caption{First passage time trajectories of neutral and charged tracers in a corrugated channel characterized by $\phi=0$ (left panel) and $\phi=\pi$ (right panel) \PM{where $\phi$ is defined in Eq.~\ref{eq:channel}}.}
 \label{fig:channel}
\end{figure}

In this contribution we study the MFPT of both charged and neutral tracers across channels of varying cross-section characterized by charged channel walls. Our results show a remarkable dependence of the MFPT on particle charge as well as on channel corrugation. For positively charged channel walls, positive (negative) tracers are depleted (attracted) towards the channel walls and their MFPT is enlarged (reduced). We show that this asymmetric response is especially enhanced when the Debye length, $\kappa^{-1}$ is comparable to the channel average section $h_0$, consistent with previous results on asymmetric charged tracer motion through inhomogeneous channels~\cite{letter-electrokin,paolo_macromolecules}. Moreover, the MFPT is sensitive to the direction in which the channel is crossed.
In particular, such a feature persists also for neutral tracers hence underlying its geometrical origin.
We exploit a systematic procedure to  approximate the tracer MFTP and develop  a framework that allows us to disentangle the geometric (entropic) contribution from the electrostatic (enthalpic) and therefore to identify the interplay between the geometrical constraints and the inhomogeneous distribution provided by the electrostatic interactions.

The structure of the text is the following: in section II we derive the $1D$ effective equation for charged tracers moving in a channel of varying cross-section, in section III we present our results and in section IV we summarize our conclusions.

\section{II Theoretical Framework}

To capture the main features of the interplay between the geometrically induced local rectification provided by the varying-section\PM{of} the channel and the electrostatic field, we study a $z-z$ electrolyte embedded in a \PM{channel of varying cross-section, see Fig.~\ref{fig:channel}}.
We assume that particles are constrained in a \PM{channel of varying cross-section} whose $y$-section changes solely along the $x$-direction and it is constant along $z$. The channel section accessible to the center of mass of a point-like tracer is $2 h(x)L_z$, being $h(x)$ the half-width of the channel along the $y$-direction and $L_z$ the (constant) width of the channel along the $z$-direction.

The motion of a suspension of non-interacting charged particles is characterized by a convection-diffusion equation, which in the overdamped regime reads
\begin{equation}
\frac{\partial P_q({\bf x},t)}{\partial t}=D\beta\nabla\cdot\left(eqP_q({\bf x},t)\nabla W({\bf x})\right)+D\nabla^{2}P_q({\bf x},t)
\label{eq:adv-diff}
\end{equation}
where  $P_q$ quantifies the probability distribution of  tracers of charge $e q$, where $e$ stands for the elementary charge, $D$ is the tracer diffusion coefficient, $\beta=1/k_BT$ the inverse thermal energy for a system at temperature $T$ (being $k_B$ the Boltzmann constant) and $W({\bf x})$ is the total conservative potential acting on the tracers. 
We encode the presence of the channel and the electrostatic potential, in the overall potential $W({\bf x})$ defined as: 
\begin{eqnarray}
 W({\bf x})=\left\{\begin{array}{cc}\psi({\bf x}), &  |y|\le h(x)\, \&\, |z| \le L_z/2 \\
 \infty, &  |y|> h(x)\, \mbox{or}\, |z|>L_z/2 \\
 \end{array}
 \right.
\label{potential}
\end{eqnarray}
that is periodic along the longitudinal direction $x$, $W({\bf x})= W({\bf x}+L{\bf e}_x)$, constant along the $z$ direction and confines particles inside the channel. 
In order to find the electrostatic potential, $\psi({\bf x})$, inside the channel, we should solve the Poisson equation
\begin{figure}
 \includegraphics[scale=0.5]{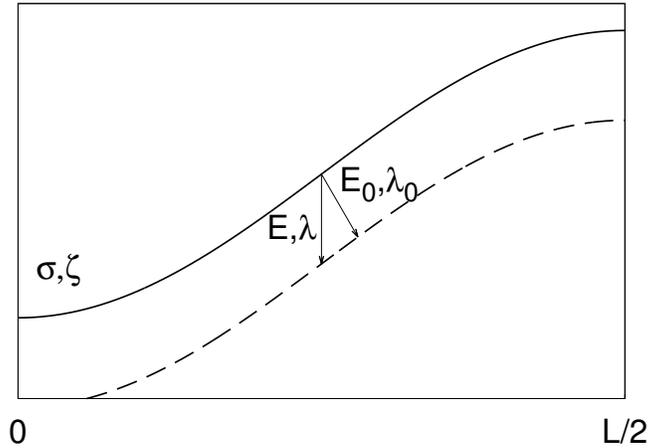}
 \caption{Debye double layer inside a \PM{channel of varying cross-section}. The Debye length $\kappa_0^{-1}=\lambda_0$ is shown as well as the approximated Debye length $\kappa^{-1}=\lambda$.}
 \label{channel-zoom}
\end{figure}
\begin{equation}
\frac{ \partial^2 \psi({\bf x})}{\partial x^2}+ \frac{ \partial^2 \psi({\bf x})}{\partial y^2}=-\frac{\rho_q({\bf x})}{\epsilon}
 \label{poisson}
\end{equation}
where $\epsilon$ corresponds to the medium dielectric constant. The electrostatic potential has to satisfy  the boundary condition of constant potential $\zeta$ (or prescribed charge density) for conducting (or insulating) channel walls. In Eq.~(\ref{poisson}) $\rho_q=\rho_0\exp\left(-\beta ze \psi({\bf x})\right)$ corresponds to the equilibrium ion charge density inside the channel in the absence of tracers. Assuming smoothly-varying channel walls,  $\partial_x h \ll 1$, we can take advantage of the lubrication approximation, $\partial^2_x\psi({\bf x})\ll \partial^2_y\psi({\bf x})$ and  reduce Eq.~\ref{poisson} to a $1D$ equation for the potential $\psi({\bf x})$.  Since the electrostatic field is perpendicular to the channel walls, for \PM{channel of varying cross-section} we must consider  the projection of the electrostatic field along the channel  when solving the Poisson equation, as shown in Fig.~\ref{channel-zoom}. For smoothly-varying amplitude channels, for which $d_x h(x)\ll 1$, the projected electrostatic field reduces to
\begin{equation}
 E=E_0\cos(\alpha)=E_0\left[1-\frac{1}{2}\left(\frac{d h(x)}{d x} \right)^2\right],
\end{equation}
where $\alpha=\arctan\left(d_x h(x)\right)$. Therefore, this geometric correction is of second order in $d_x h(x)$ and can be neglected in the following.

We can exploit the regime $d_x h(x) \ll 1$, for which the lubrication approximation holds, to simplify Eq.~\ref{eq:adv-diff}. Specifically, we factorize $P_q({\bf x}, t)$ to arrive at
\begin{eqnarray}
&P_q({\bf x},t)  =  p_q(x,t)\frac{e^{-\beta eq\psi({\bf x})}}{e^{-\beta A_q(x)}} \\
&e^{-\beta A_q(x)}  =  \dfrac{1}{2L_z h_0}\int_{-L_z/2}^{L_z/2}\int_{-h(x)}^{h(x)}e^{-\beta eq\psi({\bf x})}dy dz,
\label{free-en}
\end{eqnarray}
where $h_0$ is the average amplitude of the channel. Eq.~\ref{free-en} reproduces the well-known Fick-Jacobs approximation~\cite{Zwanzig,Reguera2001,Reguera2006,Kalinay2008,Schimansky2013} that has been exploited in diverse situations, such as entropic resonance~\cite{Entr_resonance}, cooperative rectification~\cite{paolo_pre_2012} and entropic splitters~\cite{Entr_splitter}.  
Integrating over the channel cross section, we   obtain
\begin{equation}
 \frac{\partial p_q(x,t)}{\partial t}=\frac{\partial}{\partial_x} D\left[ \beta p_q(x,t)\frac{\partial A_q(x)}{\partial x}+\frac{ \partial p_q(x,t)}{\partial x}\right]
\label{FJ1}
\end{equation}
where now the confinement is encoded in the effective potential 
\begin{equation}
A_q(x)=-\beta^{-1}\ln\left [\dfrac{1}{2h_0}\int_{-h(x)}^{h(x)}e^{-\beta eq \psi(x,y)}dy\right]
\label{free-energy}
\end{equation}
that, being the integral  of the Boltzmann weight over  all possible configurations for a given longitudinal position, $x$, can  be interpreted as an equilibrium free energy. In the last step we have taken advantage of the fact that all the quantities of interest are independent of $z$. Hence, without loss of generality we have assumed $\int_{-L_z/2}^{L_z/2}dz=1$. Defining the average, $x-$dependent, electrostatic energy as
\begin{equation}
 \langle U_q(x)\rangle=e^{\beta A_q(x)}\dfrac{1}{2h_0}\int_{-h(x)}^{h(x)}eq\psi(x,y)e^{-\beta eq \psi(x,y)}dy
\label{avg-V}
\end{equation}
from Eq.~\ref{free-en} we can define the entropy along the channel as $TS_q(x)=\langle U_q(x)\rangle -A_q(x)$ hence getting: 
\begin{equation}
 S_q(x)=\ln\left[\dfrac{1}{2h_0}\int_{-h(x)}^{h(x)}e^{-\beta eq \psi(x,y)}dy\right]+\beta \langle U_q(x)\rangle.
\label{entropy}
\end{equation}

In order to keep  analytical insight, we assume low salt concentration in the electrolyte and a small potential on the channel walls,  $\zeta$, i.e. $\beta e \zeta \ll 1$. In this regime we can linearize the Poisson-Boltzmann equation, hence the electrostatic potential inside a conducting walls channel (similar results can be obtained for insulating channel walls) reads
\begin{equation}
 \psi(x,y)=\zeta\frac{ \cosh(\kappa y)}{\cosh(\kappa h(x))},
 \label{eq:psi-def}
\end{equation}
where $\kappa=\sqrt{4 \pi \ell_B z^2 (\rho_+(x)+\rho_-(x))}$ is the inverse Debye length for an electrolyte of valence $z$ and solution ionic strength $\rho_0z^2$, $\ell_B=\beta e^2 /4\pi\epsilon$ stands for the electrolyte Bjerrum length. In this linear regime, when also $\beta eq \psi(x,y) \ll 1$, we can linearize Eq.~\ref{entropy} getting
\begin{equation}
 S_0(x)\simeq \ln \dfrac{2 h(x)}{2h_0},
\end{equation}
where the entropy has a clear geometric interpretation, being the logarithm of the space, $2h(x)$, accessible to the center of mass of a point-like tracer. Accordingly,  we introduce the entropy barrier for neutral tracers, $\Delta S_0$, defined as
\begin{equation}
 \Delta S_0=\ln \frac{h_{max}}{h_{min}},
\label{entropy-barrier}
\end{equation} 
which represents the difference, in the entropic potential, evaluated at the maximum, $h_{max}$, and minimum, $h_{min}$ of the channel aperture. Finally, we can define the total effective  free energy difference as
\begin{equation}
 \Delta A_q=A_q(h_{max})-A_q(h_{min})=\Delta\langle V_q\rangle -T\Delta S_q,
\label{delta-A}
\end{equation}
which using Eqs.~\ref{avg-V},\ref{entropy} leads to
\begin{equation}
 \Delta A_q=-\beta^{-1}\ln\left[\frac{\int_{-h_{max}}^{h_{max}}e^{-\beta eq \psi(x_M,y)}dy}{\int_{-h_{min}}^{h_{min}}e^{-\beta eq \psi(x_m,y)}dy}\right]
\label{delta-A2}
\end{equation}
 
\section{III Results}
We will analyze the motion of charged tracers in a channel whose half section along the $y$-direction is characterized by 
\begin{equation}
 h(x)=h_0-h_1 \cos \left(\frac{2\pi x}{L}+\phi\right)
\label{eq:channel}
\end{equation}
where $h_0$ is the average channel section, and $h_1$ is its modulation amplitude and assume the channel to be flat along the $z$-direction. $\phi$ controls the channel shape with respect to its boundaries fixed at $x=0$ and $x=L$. Accordingly, the maximum and minimum channel apertures read $h_{max}=h_0+h_1$ and $h_{min}=h_0-h_1$, respectively.

In order to characterize the diffusion of tracers in such channels, we study the time tracers take to get at a prescribed channel end for the first time. In particular, we focus on the mean of such a quantity, namely the mean first passage time (MFPT) tracers take to pass across the channel. 
\begin{figure*}
\includegraphics[scale=0.45]{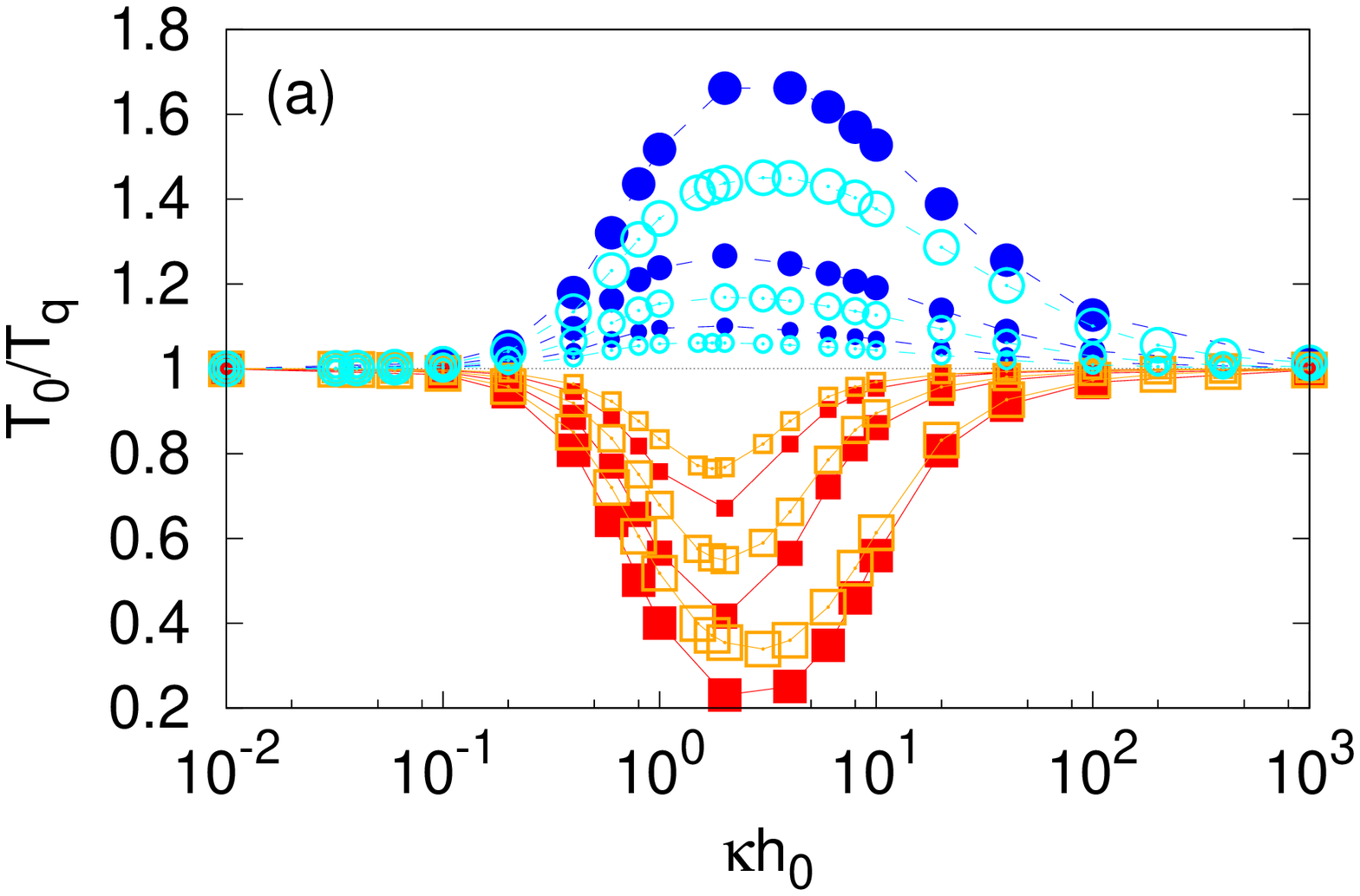}\includegraphics[scale=0.45]{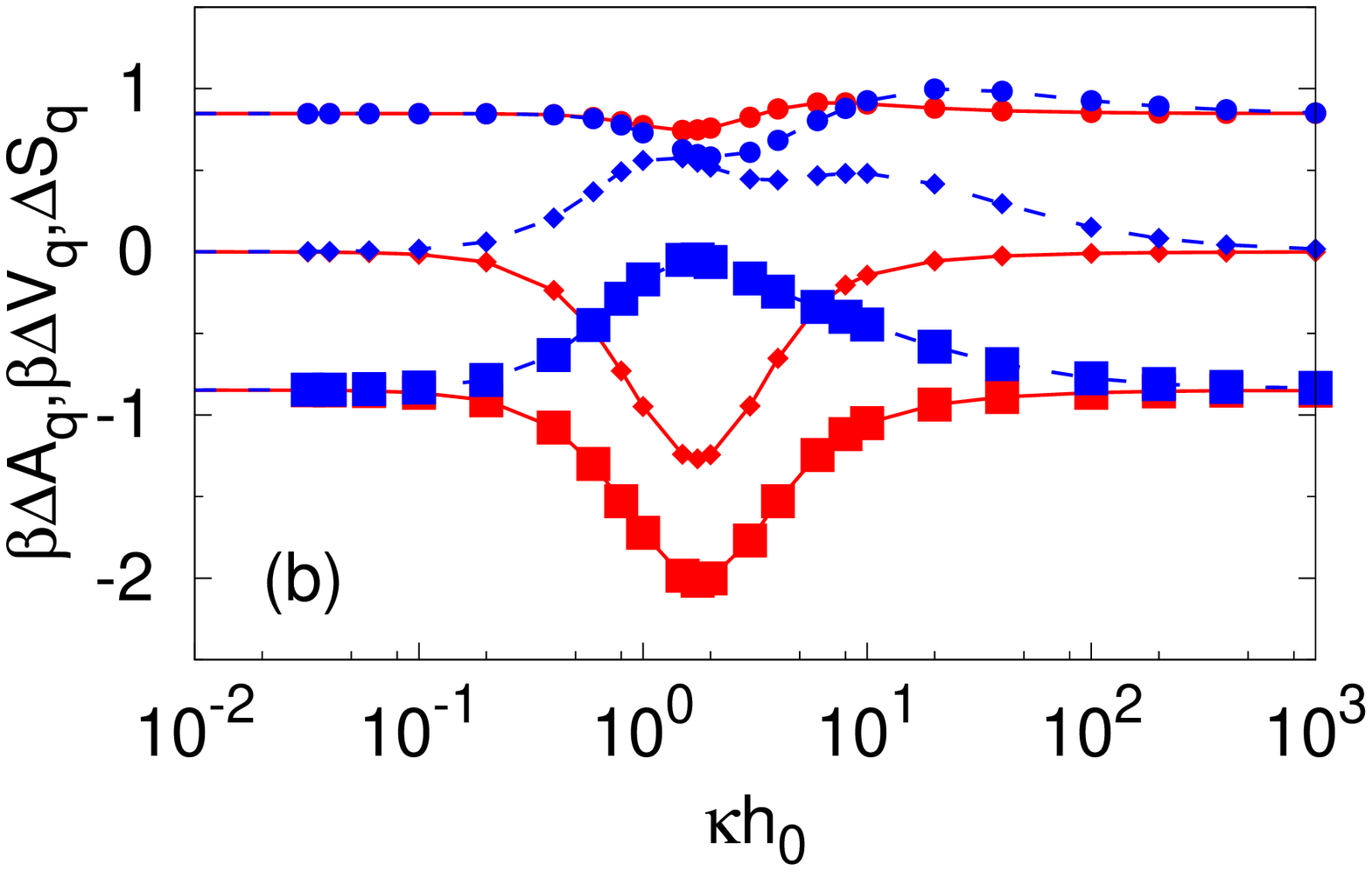}
\includegraphics[scale=0.45]{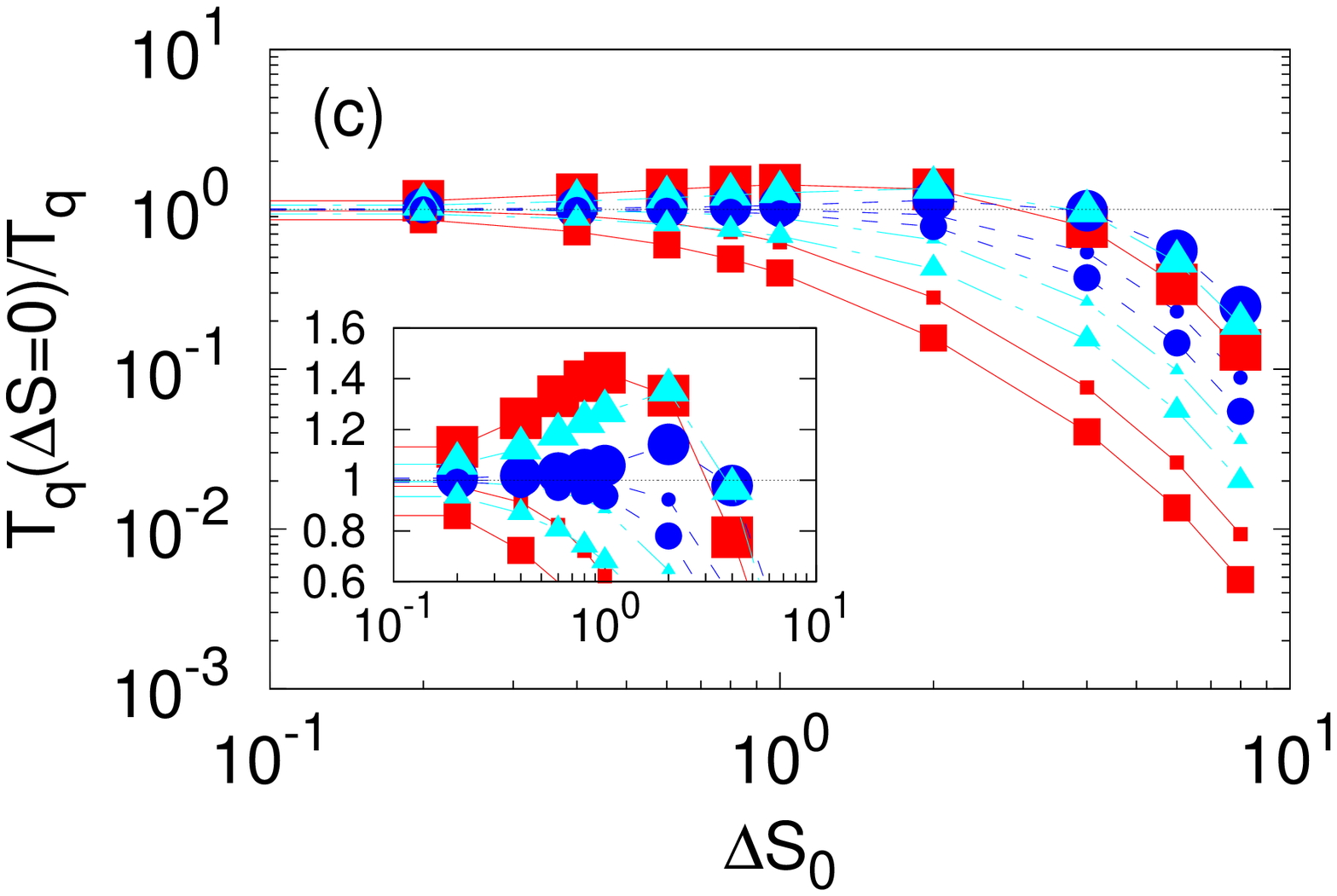}\includegraphics[scale=0.45]{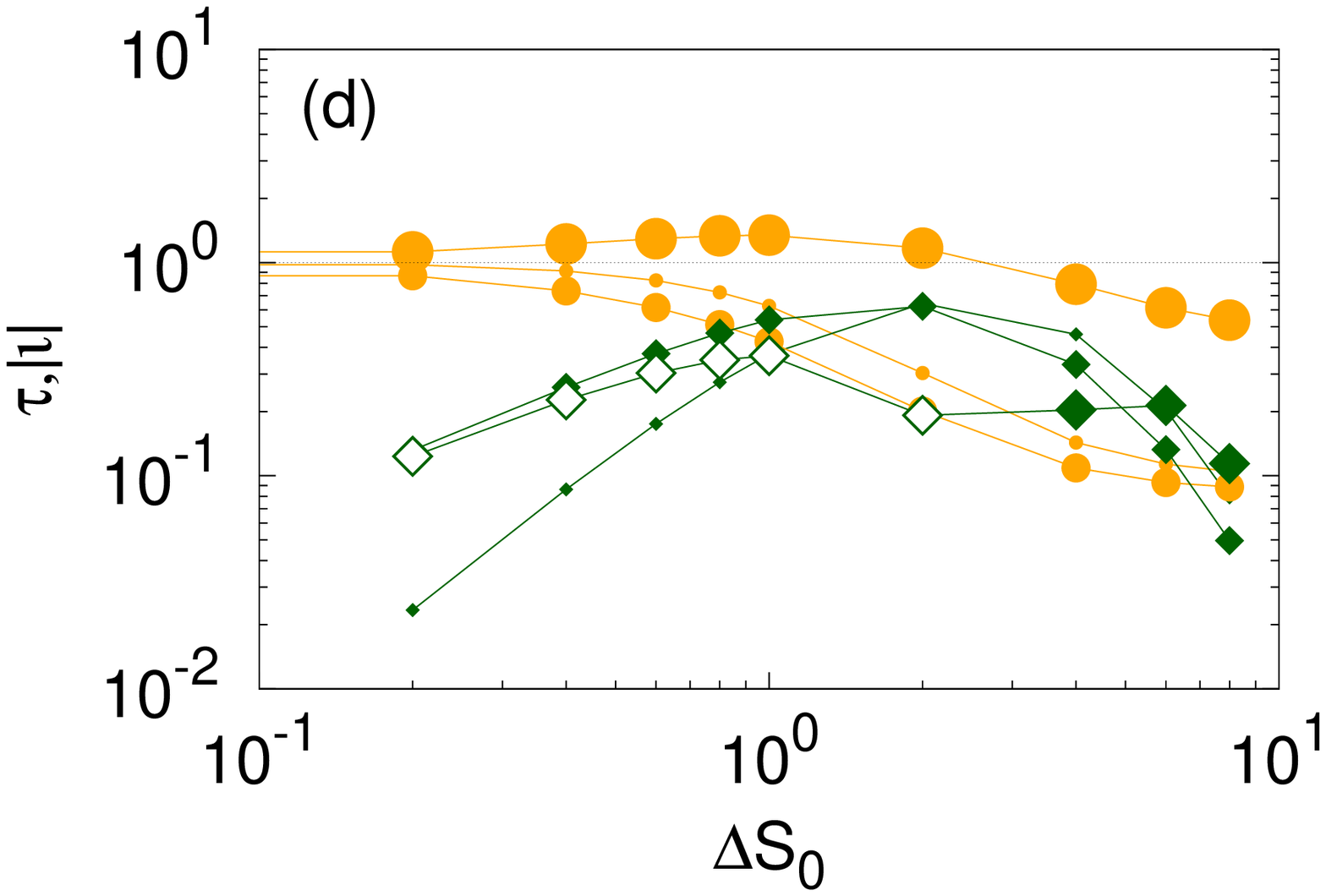}
\caption{\PM{(a)}: Filled points: inverse of the MFPT, $1/T_{q}$, \PM{obtained from the numerical solution of Eq.~\ref{eq:MFPT} and} normalized by the MFPT of neutral tracers $T_0$, as a function of the inverse Debye length, $\kappa^{-1}$, normalized by the average channel amplitude $h_0$ for positive, $q=3$, (red squares) or negative, $q=-3$, (blue dots) tracers in a conducting channel characterized by $\beta e \zeta=1$ and $\Delta S_0=0.84,1.4,2.2$ where bigger points stands for larger values of $\Delta S_0$. Open points: theoretical prediction of the MFPT provided by Eq.~\ref{eq:MFPT-simpl}. \PM{(b)}: total free energy difference $\beta \Delta A_{q}$ (squares), defined in Eq.~\ref{free-en}, entropy barrier, $\Delta S_{q}$ (circles) defined in Eq.~\ref{entropy} and enthalpic barrier, $\beta \Delta V_{q}$ (diamonds) defined in Eq.~\ref{avg-V} as a function of the inverse Debye length, $\kappa^{-1}$ for positive, $q = 3$ (red solid lines), and negative,  $q = -3$ (blue dashed lines) tracers, with $\Delta S_0=0.51$.
\PM{(c)}: inverse of the MFPT \PM{obtained from the numerical solution of Eq.~\ref{eq:MFPT} and} normalized by the MFPT for $\Delta S_0=0$, as a function of the entropic barrier $\Delta S_0$ for positive (red squares), $q=3$, or negative (blue dots), $q =-3$, tracers with $\kappa h_{0}=1$, $\beta e \zeta=1$, for $\phi=0,\pi/2,3\pi/2$ standing bigger points for larger values of $\phi$. \PM{(d)}: ratio of the MFPTs $\tau = T_-/T_+$ (orange circles), and current, $\iota$ (green diamonds) as a function of the entropic barrier $\Delta S_0$. \PM{MFPTs are obtained from the numerical solution of Eq.~\ref{eq:MFPT}} for the same parameters as in panel \PM{(c)}.}
\label{MFPTvsK}
\end{figure*}
In the following we assume that one of the ends of the channel, namely the one at $x=0$, is in contact with a reservoir of tracers and we measure the MFPT that positive, negative or neutral tracers, $t_q(x)$,  take to diffuse from a given position   inside the channel $x$ to the channel end  at $x=L$. Such a situation corresponds to a reflecting boundary condition on the end of the channel in contact with the reservoir, {\sl i.e } at $x=0$, and to an absorbing condition at the other end, at $x=L$~\footnote{The same results are valid also for the MFPT a tracer takes to diffuse from the center, $x=0$ to the boundaries of a channel of total length $2L$ and symmetric about $x=0$ with absorbing boundary conditions at $x=-L$ and $x=L$.}. Unless otherwise specified, we assume $\phi=0$ in Eq.~\ref{eq:channel},  for which channel bottlenecks are located at the channel ends. Taking advantage of the $1D$ projection, Eq.~\ref{FJ1}, we  calculate the $x$-dependent MFPT, $t(x)$, from~\cite{Risken}: 
\begin{equation}
\beta D\frac{d A_q(x)}{d x}\frac{d t_q(x)}{d x}+D\frac{d^2 t_q(x)}{d x^2}=-1.
\label{eq:MFPT}
\end{equation}
From the numerical solution of this expression, the MFTP of tracers is derived, $T_q=t_q(0)$.
Fig.~\ref{MFPTvsK} shows the inverse MFPT for positive and negative tracers  across a \PM{channel of varying cross-section} normalized by the MFPT of neutral tracers, $T_0$, whose MFPT does not depend on the Debye length $\kappa^{-1}$. When $\kappa^{-1}$ is comparable with the channel average amplitude, $h_{0}$, negative tracers benefit from the attraction to the positively charged and display an enhanced diffusion. On the contrary positive tracers, depleted from the walls, suffer a caging effect due to the entropic barrier resulting in a larger MFPT as compared to that of negative or neutral tracers. Such feature reminds the one observed for tracers in a porous media obtained by coarse-grained numerical simulations~\cite{ignacioEPL,ignacio_Faraday}. Interestingly, such a modulation in the MFPT for charged tracers vanishes for $\kappa h_{0}\ll1$ as well as for $\kappa h_{0}\gg 1$,  in agreement with the entropic electrokinetic regime observed for tracers under external forcing~\cite{letter-electrokin} and under   chemical potential gradients~\cite{paolo_macromolecules}.

We can gain insight in the dependence of the MFPT on  $\kappa$ by analyzing the effective barrier experienced by the tracers, quantified by the free energy difference $\Delta A_q$ as defined in Eq.~\ref{delta-A2}.
Fig.~\ref{MFPTvsK}\PM{(b)} shows that the dependence of $\Delta A_q$ on $\kappa$ is sensitive to tracers' charge. In particular, for $\kappa h_{0}\rightarrow 1$ the free energy barriers of negative tracers (blue squares) diminishes while the opposite holds for positive tracers (red squares). Such a diverse behavior of $\Delta A_q$ according to tracer charge explains the different behavior of the MFPT shown in Fig.~\ref{MFPTvsK}: positive tracers, which experience an enhanced free energy barrier, will take longer to cross the channel as compared to negative tracers. 

Using Eqs.~\ref{free-en},~\ref{avg-V},~\ref{entropy} we can separately quantify  the entropic and enthalpic contributions to the effective free energy difference. As shown in Fig.~\ref{MFPTvsK}\PM{(b)}, the entropic contribution (circles) is mildly affected by variations in $\kappa h_0$ whereas the enthalpic contribution (diamonds) shows a strong sensitivity. Moreover, while for positive tracers (red points) the enthalpic and a entropic contribution sum up amplifying the magnitude of $\Delta A_{q}$, for negative tracers (blue points) the two contributions have different sign therefore reducing the magnitude of $\Delta A_{q}$. 

We can simplify Eq.~\ref{eq:MFPT}, even in the non-linear regime $\beta eq \psi > 1$, by assuming $\partial_x A_q(x)$ to be piece-wise linear. Choosing $\partial_x A_q(x)=\pm 2\frac{\Delta A_q}{L}$ for $x<L/2$ or $x>L/2$ respectively, we can analytically solve Eq.~\ref{eq:MFPT} getting
\begin{equation}
 \bar T_q=\frac{L^2}{D\beta\Delta A_q^2}\left(\cosh\beta\Delta A_q -1\right).
\label{eq:MFPT-simpl}
\end{equation}
In the limit $\Delta A_q\rightarrow 0$, Eq.~\ref{delta-A} leads to $\bar T_q=\frac{L^2}{2D}$ in agreement with Eq.~\ref{eq:MFPT}. When $\Delta A_q\ne0$, we can substitute the values of $\Delta A_q$ derived from Eq.~\ref{delta-A} into  Eq.~\ref{eq:MFPT-simpl}. Fig.~\ref{MFPTvsK}(a) shows that this approximation provides a very good quantitative agreement with the numerical solutions. 

In the linear regime, $\beta eq \psi(x,y)\lesssim 1$, we can obtain analytic expressions for the effective free energy barrier experienced by a charged tracer
\begin{equation}
\Delta A\simeq-\beta^{-1}\left\{\ln\left[\frac{h_{max}}{h_{min}}\right]+\ln\left[1+\beta\zeta eq\Delta\Theta\right]\right\}
\label{A-exp}
\end{equation}
where
\begin{equation}
 \Delta\Theta=\frac{\tanh\left(\kappa h_{min}\right)}{\kappa h_{min}}-\frac{\tanh\left(\kappa h_{max}\right)}{\kappa h_{max}}
\end{equation}
which shows that the enthalpic contribution always vanishes  when $\beta\zeta eq\rightarrow 0$ and $\kappa h\rightarrow \infty$ or $\kappa h\rightarrow 0$, and $\Delta A_q$ reduces to the entropic barrier, $\Delta S_0$ experienced by neutral tracers.
On the contrary, when $\kappa h\simeq 1$ the enthalpic contribution is relevant and $\Delta A_q$ retains a dependence the charge: positive (negative) tracers experience an  enhanced (reduced) free energy barrier, as shown in Fig.~\ref{MFPTvsK}\PM{(b)}.

We can characterize the dependence of the MFPT on the channel geometry by considering a  prescribed electrolyte, $\kappa$, and a charged channel with electrostatic potential, $\zeta$, and vary the entropic barrier, $\Delta S_0$.  For $\phi=0$ (similar results have been obtained for $\phi=\pm\pi$) Fig.~\ref{MFPTvsK}\PM{(c)} shows a monotonous increase in the MFPT for all tracers upon increasing $\Delta S_0$. For neutral tracers such an increase in the MFPT is the signature of the entropic-induced modulation in tracer transport  due to the varying-section of the channel. In particular, positive tracers are the most sensitive to the entropic barrier whereas negative tracers are the least sensitive to variations in $\Delta S_0$, in agreement with the prediction of Eq.~\ref{A-exp}. For $\phi\neq0,\pm\pi$, Fig.~\ref{MFPTvsK}\PM{(c)} shows two opposite regimes. While for $\phi=\pi/2$ the behavior is similar to $\phi=0,\pm\pi$ except that all tracers experience an enhanced dependence on $\Delta S_0$, for $\phi=3\pi/2$, the dependence of the MFPT on $\Delta S_0$ is no longer monotonic and the MFPT is minimized for a non vanishing value of $\Delta S_0$ , in agreement with previous general models that accounted for only enthalpic contributions~\cite{Palyulin}. Moreover, in the range of values of $\Delta S_0$ for which the effective free energy barrier reduces  the MFPT, positive tracers are faster than neutral and negative ones. In contrast, for larger values of $\Delta S_0$, negative tracers are faster than neutral and positive ones. 

The asymmetric response of  positive and negative tracers to the channel corrugation can be useful to control their relative motion and positioning along a channel, with relevant applications, such as chemical segregation or particle separation. Hence,  it is interesting to quantify the ratio of the MFPT of positive and negative tracers, namely: 
\begin{equation}
 \tau=\frac{T_+}{T_-}
\end{equation}

As shown in Fig.~\ref{MFPTvsK}\PM{(d)}, for uniform channels, $\Delta S_0=0$, positive and negative tracers experience the same MFPT while for increasing $\Delta S_0$ the dependence of $\tau$ on $\Delta S_0$ is sensitive to $\phi$. For $\phi=0,\pm\pi$, negative tracers can be up to ten times faster than positive ones. On the contrary, for $\phi=3\pi/2$ the ratio between positive and negative tracers is larger than unity for smaller values of $\Delta S_0$ and eventually is smaller than unity for increasing values of $\Delta S_0$.

The asymmetric response of positive and negative tracers means that if released homogeneously, they will induce a transient electric current due to the channel inhomogeneous section. This current can be estimated using the MFTPs through the quantity 
\begin{equation}
 \iota=\frac{L^2}{2D}\frac{T_--T_+}{T_-T_+}
 \label{current}
\end{equation}
For $\phi=0,\pm\pi$, Fig.~\ref{MFPTvsK}\PM{(d)} shows a non monotonous behavior of $\iota$ on $\Delta S_0$, displaying   a maximum for $\Delta S \sim 1$. In contrast, for $\phi=3\pi/2$, the sign of the current changes with $\Delta S_0$ switching from positive currents, for smaller values of $\Delta S_0$, to negative currents for larger values of $\Delta S_0$. 

\begin{figure*}
\includegraphics[scale=0.45]{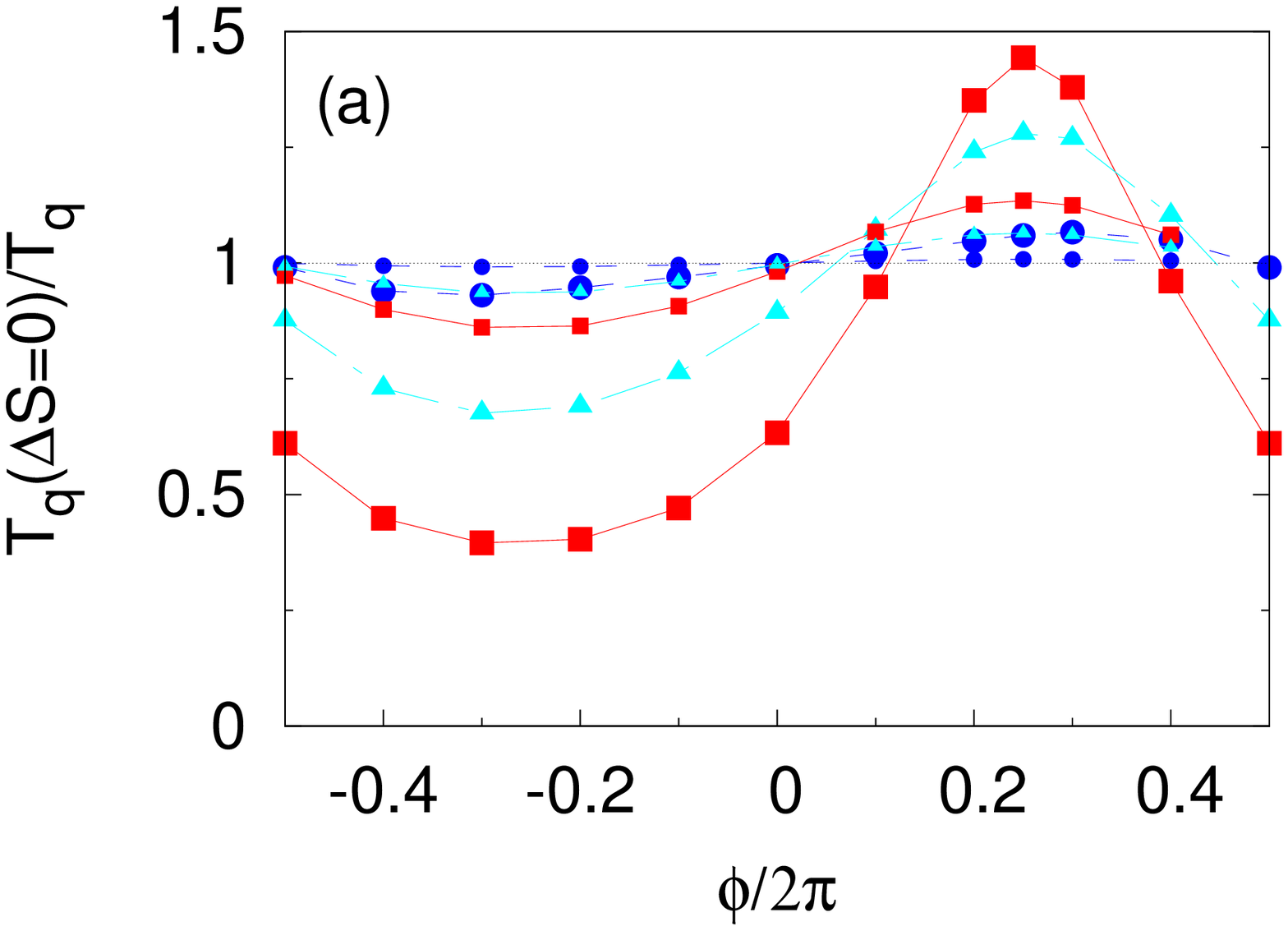}\includegraphics[scale=0.45]{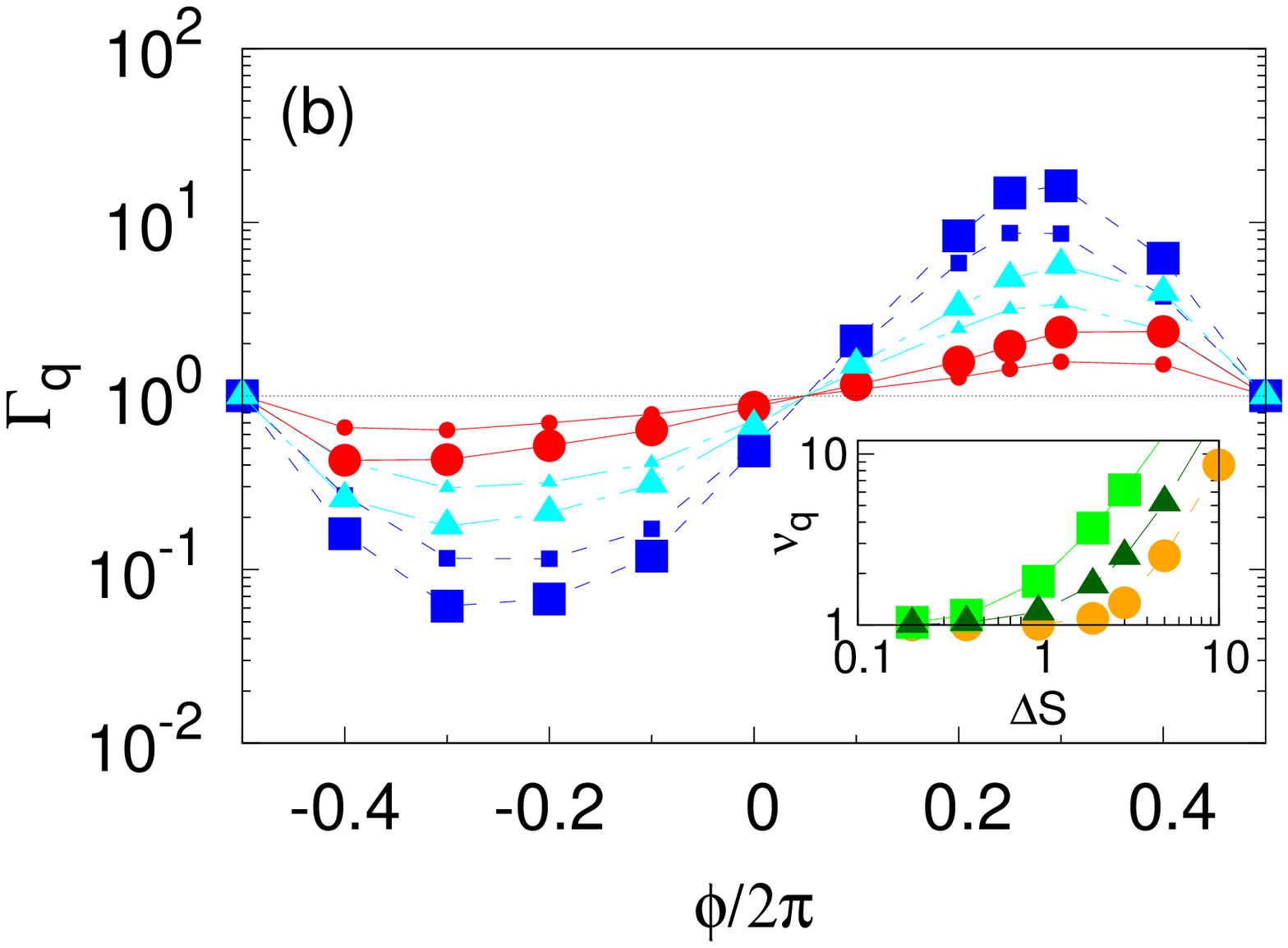}
\includegraphics[scale=0.45]{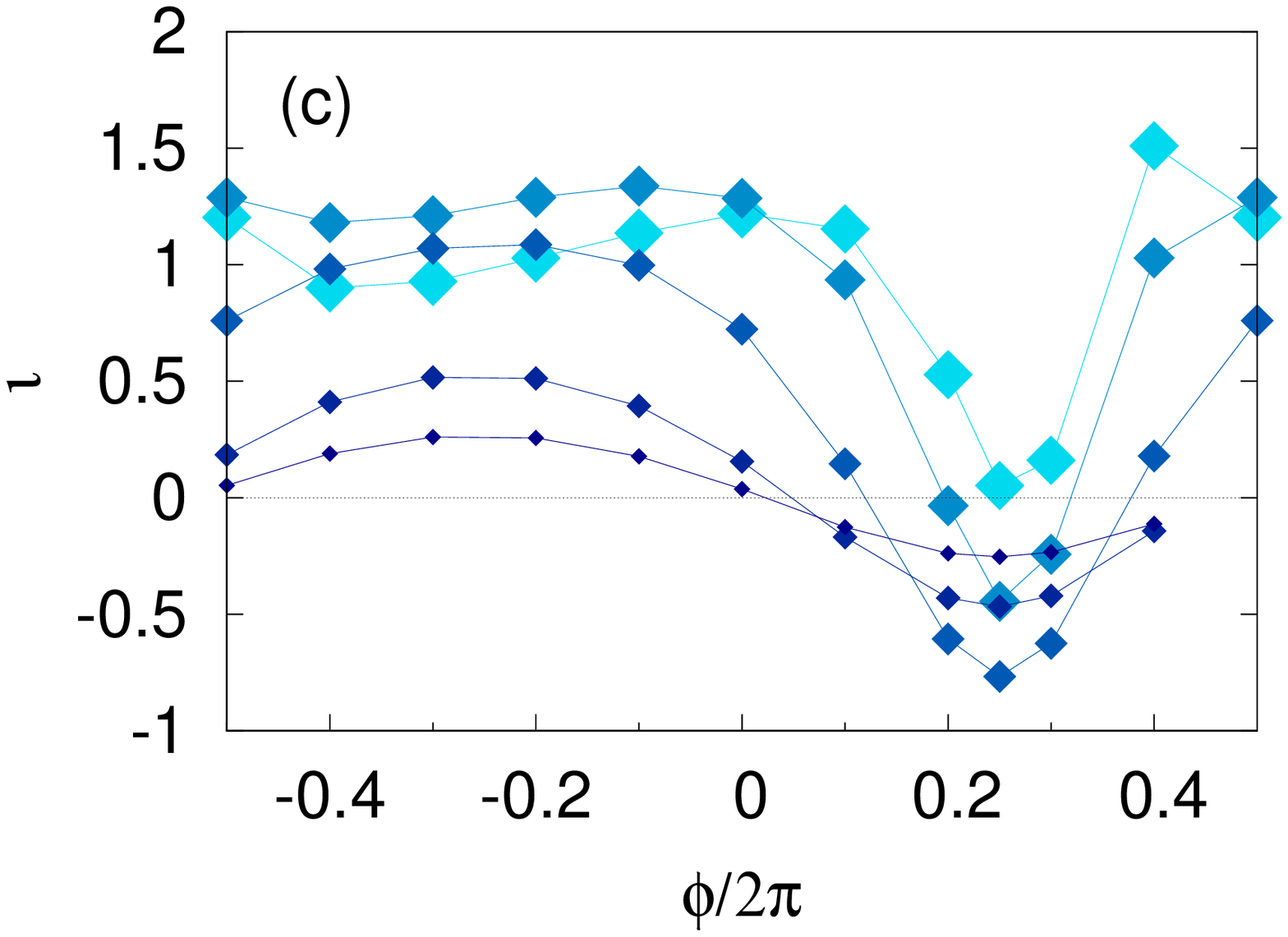}\includegraphics[scale=0.45]{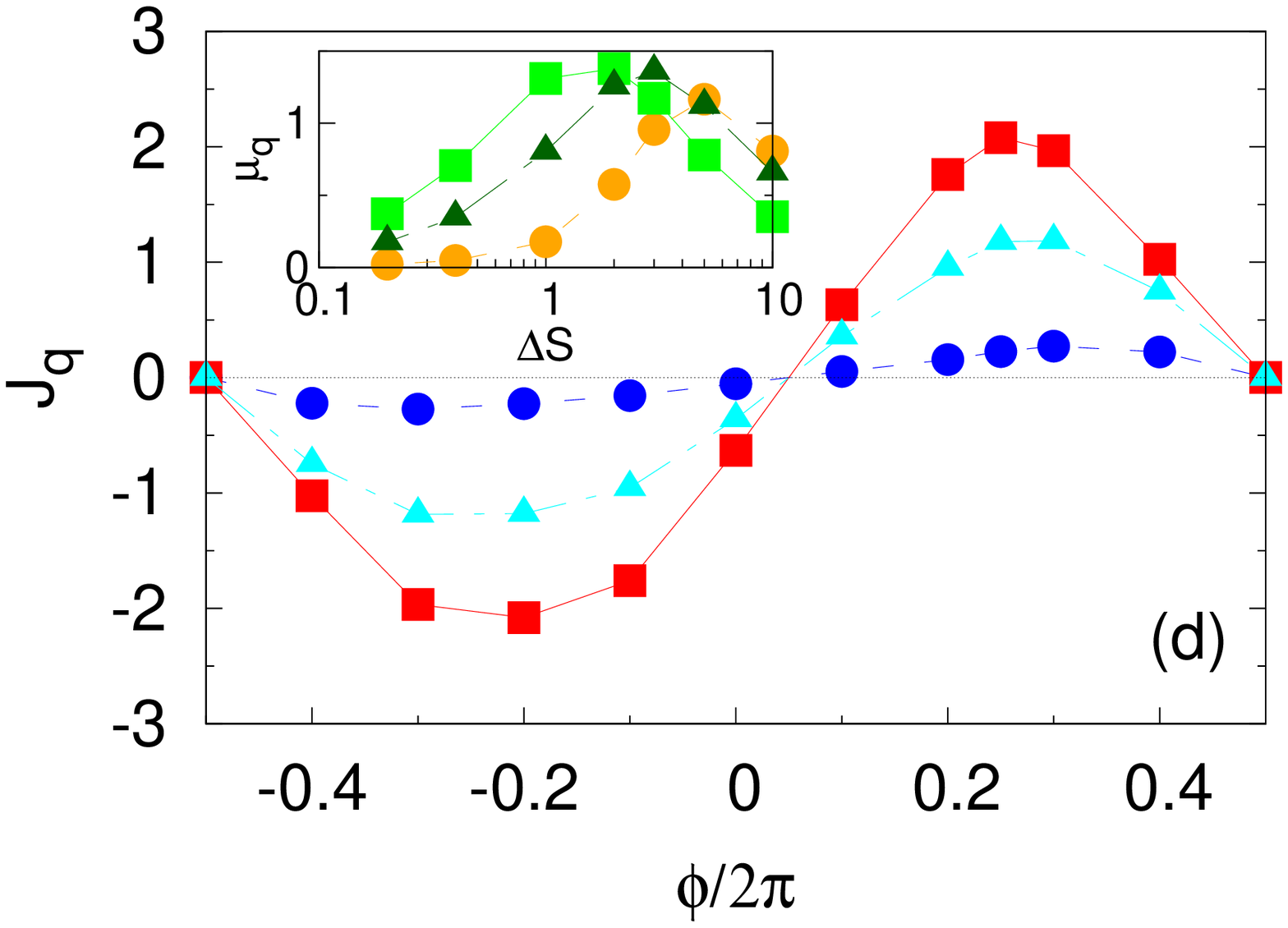}
\caption{\PM{(a)}: Inverse of the MFPT, $T_q$, \PM{obtained from the numerical solution of Eq.~\ref{eq:MFPT} and} normalized by the MFPT for $\Delta S_0=0$, as a function of the channel shape characterized by the phase $\phi$ with $\beta e \zeta=1$, $\kappa h_0=1$ and $\Delta S_0=0.2,1$ (bigger points correspond to larger values of $\Delta S_0$) for positive, $q=3$ (red squares), negative, $q=-3$ (blue dots) and neutral, $q=0$ (cyan triangles), tracers. 
\PM{(b)}: ratio of the MFPT\PM{, shown in panel (a),} of tracers diffusing along opposite directions, $\Gamma_q$, as a function of the phase, $\phi$, for positive, $q=3$ (red squares), negative, $q=-3$ (blue dots) and neutral, $q=0$ (cyan triangles), tracers, $\kappa h_0=1$ and $\Delta S_0=0.2,1$ (bigger points correspond to larger values of $\Delta S_0$). Inset: Dependence of $\nu_q$ on $\phi$ for positive (light green squares), negative (orange dots), and neutral (dark green triangles), tracers (same parameter values as in the main panel).  
\PM{(c)}: Dimensionless difference in the effective tracer currents, $\iota$ as a function of $\phi$, for different values of the entropic barrier $\Delta S_0=0.2,0.4,1,2,3$ where bigger points and darker lines correspond to larger values of $\Delta S_0$, with  $\kappa h_0=1$ and $\beta e \zeta=1$.  \PM{MFPTs are obtained from the numerical solution of Eq.~\ref{eq:MFPT}.}
\PM{(d)}: Dimensionless difference in the effective tracer currents, $\Pi_{q}$ as a function of $\phi$, for positive, $q=3$ (red squares), negative, $q=-3$ (blue dots), and neutral, $q=0$ (cyan triangles), tracers, with $\Delta S_0=1$, $\kappa h_0=1$ and $\beta e \zeta=1$. \PM{MFPTs are obtained from the numerical solution of Eq.~\ref{eq:MFPT}}. Inset: Dependence of $\mu_{q}$, for positive (light green squares), negative (orange dots) and neutral (dark green triangles) tracers (same parameter values as in main panel).
}
\label{MFPTvsPhase}
\end{figure*}

Since the MFPT displays a non-trivial dependence on the channel geometrical details~\cite{Redner}, we have systematically studied the dependence of the MFPT on $\phi$, as shown in Fig.~\ref{MFPTvsPhase}\PM{(a)}. 
In particular, Fig.~\ref{MFPTvsPhase}\PM{(a)} shows that for a large set of values of $\phi$, positive tracers take more time than neutral and negative ones to leave the channel, as depicted for $\phi=0$. As already anticipated in Fig.~\ref{MFPTvsK}\PM{(c)}, for $\phi\sim \pi/2$ and small values of the entropic barrier, $\Delta S_0\lesssim 1$, the MFPT of all tracers is smaller than the corresponding value for a flat channel, i.e. for $\Delta S_0=0$. Moreover, in this range of parameters, the MFPT of positively charged tracers is smaller than the MFPT of neutral and negative tracers. In contrast, for $\phi\nsim \pi/2$ and/or for larger values of $\Delta S_0$ tracers MFPT is smaller than the corresponding MFPT for $\Delta S_0=0$ and the MFPT of negative tracers is smaller than that of neutral and positive ones. This dependence underlines the sensitivity of the MFPT on the details of the sequence of bottlenecks and apertures. Accordingly, we expect a similar behavior for an asymmetric channel where instead of varying the phase, $\phi$, we modify the relative position of the  maximum amplitude of the channel while keeping fixed the channel bottlenecks at the boundaries. Obviously, for a set of connected channels, the dependence on $\phi$ will vanish asymptotically.


The involved dependence of the MFPT on the channel geometry  rises the question of the difference in the MFPT of a tracer along opposite directions of a prescribed channel. We can exploit  the symmetry of the channel under study and consider the ratio between the MFPT of a tracer moving in opposite directions, $\Gamma_q$, which can be obtained from the results in Fig.~\ref{MFPTvsPhase}\PM{(a)} noticing that, given $T_q(\phi)$, the MFTP along the opposite direction of the channel is $T_q(-\phi)$. Accordingly, we quantify this asymmetry with the quantity:
\begin{equation}
 \Gamma_q=\frac{T_q(-\phi)}{T_q(\phi)}
\end{equation}
that for a constant channel, $\Delta S_0=0$, leads to $\Gamma_q=1$.
Fig.~\ref{MFPTvsPhase}.B shows that positive tracers, whose MFPT is more sensitive to $\phi$, experience a remarkable dependence of $\Gamma_q$ on the phase shift, $\phi$, whereas neutral and negative tracers are less affected. Moreover, Fig.~\ref{MFPTvsPhase}\PM{(b)} shows that the modulation of $\Gamma_q$ increases with the entropic barrier, $\Delta S_0$. In order to quantify such a dependence we can look at the dispersion of the values of $\Gamma_q$, defined as:
\begin{equation}
 \nu_q=\frac{1}{2\pi}\int_{-\pi}^{\pi} \Gamma^2_qd\phi
\end{equation}
The inset of Fig.~\ref{MFPTvsPhase}.B shows a monotonic growth of $\nu_q$ upon increase of $\Delta S_0$ whereas $\nu_q\rightarrow 1$ for $\Delta S_0\rightarrow 0$.

If tracers of opposite charge are present at the same time, we can estimate the electric current generated in response to fluctuations as defined in Eq.~\ref{current}. Fig.~\ref{MFPTvsPhase}\PM{(c)}  shows the dependence of  $\iota$ on the phase $\phi$ for different values of $\Delta S_0$. Interestingly, $\iota$  is very sensitivity to the channel geometry, and its sign  and magnitude can vary significantly. As shown in Fig.~\ref{MFPTvsPhase}\PM{(c)}, for small entropic barriers, $\Delta S_0 \ll 1$ the amplitude of the deviations in the electric current profile are reduced and the profile is almost symmetric with respect to the phase $\phi$. Hence by tuning the geometry of the channel it is possible to select which tracers has the shortest MFPT. On the contrary, large values of  $\Delta S_0$ amplify the difference in the MFPT of tracers with opposite charges and the dependence of $\iota$ on $\phi$  becomes more complex. Eventually, for larger entropy barriers, $\Delta S_0\simeq 10$, the enthalpic barrier that positive tracers (depleted from channel walls) have to overcome becomes so large that their MFPT always exceed that of negative tracers, hence leading to a constant sign of $\iota$ whose amplitude still retains a dependence on $\phi$.

The asymmetry in the transport properties of the channel shown in Fig.~\ref{MFPTvsPhase}\PM{(b)} resemble that of a diode for which the magnitude of the flux varies upon inverting the boundary conditions. 
From the MFTP corresponding to each set of boundary conditions we can estimate a rectifying flux
\begin{equation}
 \Pi_q=\frac{L^2}{D}\left(\frac{T_q(-\phi)-T_q(\phi)}{T_q(\phi)T_q(-\phi)}\right),
\end{equation}
which vanishes for symmetric channels, for which $T_q(-\phi)=T_q(\phi)$. 
Fig.~\ref{MFPTvsPhase}\PM{(d)} shows the dependence of $\Pi_q$ on the phase $\phi$. In particular, $\Pi_q\neq 0$ for $\phi\neq 0,\pm\pi$, therefore the asymmetry in the MFPT 
identifies a direction along which particle can diffuse faster. The non vanishing values of $\Pi_0$ for neutral tracers, shows that such an effect has an entropic origin since for neutral tracers there is no enthalpic contribution.
The amplitude of the  entropic barrier, $\Delta S_0$ strongly affects the values of $\Pi_q$. In order to quantify such a dependence it is insightful to look at:
\begin{equation}
 \mu_q = \frac{1}{2\pi}\int_{-\pi}^{\pi} \Pi^2_q d\phi
\end{equation}
that, since $\int_{-\pi}^{\pi} \Pi_q d\phi=0$, captures the overall departure of $\Pi_q$ from the symmetric case characterized by $\Pi_q =0$. The inset of Fig.~\ref{MFPTvsPhase}\PM{(d)} shows that $\mu_q$ has non-monotonous dependence on $\Delta S_0$ and we can identify a, charge-dependent, optimal value of $\Delta S_0$ that maximizes $\Pi_q$.

\section{IV Conclusions}

We have studied the mean first passage time (MFPT) of charged and neutral tracers suspended in an electrolyte confined between  charged walls.
Our data show a remarkable dependence of the MFPT of both charged and neutral tracers on the channel geometry  when the double layer is comparable to the channel section. We have found that the MFPT depends on both the amplitude of the channel corrugation, which we quantify through an entropic parameter $\Delta S_0$ (see Fig.~\ref{MFPTvsK}\PM{(c)})  as well as on the details of the geometry of the channel captured by $\phi$, (see Fig.~\ref{MFPTvsPhase}\PM{(a)}). 
In particular, we have found a strong asymmetry in the dependence of the MFTP on $\phi$ (see Fig.~\ref{MFPTvsPhase}\PM{(b)}). 
In particular, by quantifying the difference in the MFPT for tracers diffusing along opposite directions through the corrugated channel, $\Pi_q$ (see Fig.~\ref{MFPTvsPhase}\PM{(d)}) we have found a non monotonic behavior of $\Pi_q$ upon increasing $\Delta S_0$. For $\Delta S_0\rightarrow 0$ or $\Delta S_0\rightarrow \infty$ we have $\Pi_q\rightarrow 0$ whereas $\Pi_q$ experience a maximum for $\Delta S_0\simeq 1$ (see inset of Fig.~\ref{MFPTvsPhase}\PM{(d)}). 
Moreover, for mild values of $\Delta S_0$ and for $\phi\simeq \pi/2$, we found a reduction of the MFPT of all tracers as compared to the corresponding values obtained for $\Delta S_0=0$ (see Fig.~\ref{MFPTvsK}\PM{(c)}). Therefore, the corrugation of the channel can reduce the time that the tracers needs to cross it, showing the relevance of entropic constraints as a complementary mechanism to the enthalpic forces identified earlier~\cite{Palyulin}. Interestingly, these features persist also for neutral tracers, hence underlying the entropic origin of these effects.

For charged tracers, the electrostatic interactions with the charged channel walls provide additional parameters to control the MFPT, namely their positive (negative) charge and the properties of the electrolyte, captured by the Debye length $\kappa^{-1}$. In fact, according to the sign of their charge, tracers are depleted or attracted to the channel walls hence experiencing different enthalpic barriers. 
When the Debye length, $\kappa^{-1}$ and the channel average section $h_0$ are not commensurate, namely $\kappa h_0\ll 1$ or  $\kappa h_0\gg 1$, the MFPT is quite insensitive to the charge of the tracers, similarly to what has been observed in driven systems~\cite{letter-electrokin,paolo_macromolecules}.

In the present framework we have dealt with point-like particles but the extension to finite-size particles is straightforward. As it has been discussed earlier~\cite{Entr_splitter,paolo_jcp_2013}, it is possible to incorporate the dependence on particle size taking into account that the effective width available to the tracer is reduced by its own size. Accordingly, the formalism developed can be straightforwardly extended changing the integration limits both in  the effective free energy barrier, $\Delta A$, and channel opening, $\Delta S_0$, to $h_{min}\rightarrow h_{min}-R$ and $h_{max}\rightarrow h_{max}-R$. Such a generalization clarifies the impact that tracer size and entropic constraints have on tracer MFTP: 
larger particles will experience a larger effective entropic barrier as compared to smaller ones, therefore opening a new route for tuning particle MFPT across corrugated channels.

\PM{Finally we remark that our results can be used to deduce the shape of a pore by using data coming from resistive pulse sensing experiments~\cite{Saleh2003,Ito2004,Heins2005,Arjmandi2012,Willmott2015}}. 

\section*{Acknowledgments}
P.M. acknowledges Gleb Oshanin for useful discussions.
We acknowledge MINECO and DURSI  for financial support under projects  FIS\ 2011-22603 and 2014SGR-922, respectively. I.P. and J.M.R. acknowledge financial support from {\sl Generalitat de Catalunya } under Program {\sl Icrea Acad\`emia}.

\bibliography{letter_electrokinetics}

\end{document}